\documentclass[aps,showpacs,showkeys,twocolumn,preprintnumbers,amsmath,amssymb]{revtex4}

\UseRawInputEncoding
\usepackage{hyperref}
\hypersetup{colorlinks=true,
            linkcolor=magenta,
            filecolor=black,
            urlcolor=magenta,
            citecolor=magenta
           }

\usepackage{dcolumn}

\usepackage{float}
\usepackage{graphicx}

\usepackage{epstopdf}
\usepackage{graphicx}
\usepackage{epstopdf}
\usepackage{color}
\usepackage{amsmath,amssymb,amsfonts}

\makeatletter

\newcommand{\Rmnum}[1]{\expandafter\@slowromancap\romannumeral #1@}
\makeatother

\begin{document}
\baselineskip=0.5 cm

\title[Article Title]{Circular polarization images of Sgr A* under different magnetic field geometries}

\author{Hao Yin$^{1}$, Songbai Chen$^{1,2}$\footnote{Corresponding author: csb3752@hunnu.edu.cn}, Jiliang Jing$^{1,2}$\footnote{jljing@hunnu.edu.cn}}

\affiliation{$^1$Department of Physics, Institute of Interdisciplinary Studies, Hunan Research Center of the Basic Discipline for Quantum Effects and Quantum Technologies, Key Laboratory of Low Dimensional Quantum Structures and Quantum Control of Ministry of Education, Synergetic Innovation Center for Quantum Effects and Applications, Hunan Normal University,  Changsha, Hunan 410081, People's Republic of China
\\ $ ^2$Center for Gravitation and Cosmology, College of Physical Science and Technology, Yangzhou University, Yangzhou 225009, People's Republic of China}
\begin{abstract}
\baselineskip=0.4 cm
Sgr A* exhibits a persistent negative circular polarization (CP) at 230\,GHz, offering a powerful probe of the magnetic field geometry in its accretion flow. Using a stationary semi-analytic radiatively inefficient accretion flow (RIAF) model in Kerr spacetime with polarized radiative transfer, we systematically analyze CP images for six poloidal magnetic field configurations across varying black hole spins, inclinations, and field polarities. We find that the CP production is dominated by Faraday conversion in radial, parabolic, quadrupole and combined magnetic field geometries, while it is dominated by intrinsic emission in dipole and vertical magnetic field configurations.
The radial and parabolic configurations produce the polarity-invariant net CP, while dipole and vertical fields yield the polarity-sensitive one. As the accretion disk is prograde with respect to the black hole spin, the CP production across all six field geometries is found to be lower at high spin case, while the situation is more complicated in the retrograde case.  Moreover, the net CP observed from edge-on views $V_{\rm net} \approx 0$ except for the quadrupole geometry. Comparing with ALMA data, the reversed-field model is excluded at high inclinations and then the magnetic field geometry of Sgr A* is constrained.
\end{abstract}

\maketitle
\section{Introduction}\label{sec:intro}
Recent 1.3 mm very-long-baseline interferometry (VLBI) observations of the supermassive black holes M87* and Sagittarius A* (Sgr A*) by the Event Horizon Telescope (EHT) have opened a new era for probing plasma dynamics and magnetic field structures on event-horizon scales \cite{EHTM87-I,EHTSgrA-I,EHTM87-VII,EHTSgrA-VII}. In these radiatively inefficient accretion flow (RIAF) systems, near-horizon polarimetric images carry more information than total-intensity images and impose stronger constraints on models of the accretion flow and jet-launching region \cite{EHTM87-VII,EHTSgrA-V,Chen:2022scf,ccj:polarizedimages,Qin:2025zdv,Chen:2023wna}. Among the various components of polarized radiation, circular polarization (CP), although typically only at the percent level, is sensitive to plasma properties and to the underlying magnetic field structure, making it an essential diagnostic \cite{EHTM87-IX,Ricarte:2021cpmag}.

CP can effectively reveal the magnetic structure of accretion flows and the geometric properties of magnetic fields in general relativistic magnetohydrodynamic (GRMHD) models under suitable viewing angles, optical depths, and Faraday depths \cite{Tsunetoe:2021toroidal,Moscibrodzka:2021unraveling,Ricarte:2021cpmag}. In addition,  CP is an important probe of plasma composition because CP emission and its Faraday rotation are sensitive to the electron-positron pair content \cite{Wilson:1997pair,Anantua:2020composition,Emami:2021jetcomposition}.

For Sgr A*, this issue is especially interesting because, over the past few decades, its unresolved fractional circular polarization ($V_{\rm net}$) at 1.3 mm (and at longer wavelengths) has remained surprisingly stable at negative values, consistently measured between roughly $-0.92\%$ and $-1.5\%$ \cite{Bower:2018interferometric,Wielgus:2022alma}. The persistent handedness and magnitude of the CP signal in Sgr A* strongly suggest the presence of a stable large-scale magnetic field structure and a preferred magnetic polarity \cite{Beckert:2002jets}.

Currently, first-principles GRMHD simulations are the standard theoretical framework for interpreting EHT polarization data \cite{EHTM87-VII,EHTM87-VIII,Wong:2022patoka}. However, GRMHD models face several limitations. They are not fully adequate for describing collisionless plasmas and rely on highly uncertain initial conditions, while most studies are carried out within the ideal-MHD framework, despite the recognized importance of non-ideal effects such as magnetic reconnection in turbulent accretion flows \cite{Ripperda:2022flares}. Moreover, because the magnetic field and plasma coevolve self-consistently in ideal GRMHD, it is difficult to isolate and test the impact of specific magnetic field geometries on observable properties. Global stationary semi-analytic RIAF models provide a complementary approach \cite{Saurabh:2025m87semi,Wang:2025btn,Long:2025nwk}. Unlike computationally expensive GRMHD simulations with implicitly coupled parameters, semi-analytic models allow parametric studies of the accretion flow and enable us to investigate how disk properties and magnetic field geometries affect CP signals. 

RIAF models feature a thick, nearly virialized disk structure with 
plasma properties parameterized by radial power laws and vertical Gaussian profiles. The flow is assumed to orbit azimuthally outside the ISCO with angular velocity as a fixed fraction of the Keplerian value, while fluid inside the ISCO  plunge radially inward on constant angular momentum trajectories. The vertical scale height is prescribed as a fraction of cylindrical radius, allowing simulation 
of different disk thicknesses. Although these semi-analytical models involve several physical simplifications, including the assumptions on stationarity and the power-law structure of the flows, they capture the fundamental characteristics of radiatively inefficient accretion flows.  These advantages of RIAF models enable efficient parameter space exploration for mm-VLBI imaging \cite{Broderick:2011lowspin,Broderick:2005jj}.

RIAF models have successfully been applied to various aspects: interpreting multi-frequency models of black hole photon rings \cite{Desire:2024mzp}, as well as testing the non-Kerr spacetime features \cite{Vincent:2020dij,Yin:2025rao,Zhang:2024jrw,Zeng:2025kyv,Zeng:2026ntc,Liu:2025wwq}. Recent studies further demonstrate that RIAF models can directly fit to EHT data \cite{SaraerToosi:2025mvl}. In polarization studies,  RIAF models can reproduce the linear polarization fraction and EVPA modes observed by the EHT \cite{Saurabh:2025m87semi}. However, CP images of black hole accretion flows are less well studied than their linearly polarized counterparts.

In this paper, we use a stationary semi-analytic RIAF model in Kerr spacetime to systematically study the CP images of Sgr A* for different magnetic field geometries, focusing on the net circular polarization fraction ($V_{\rm net}$). We explore models spanning different black hole spins, observer inclinations, and both aligned and reversed global magnetic field polarities in order to understand how $V_{\rm net}$ depends on magnetic field geometry. Finally, we compare our synthetic models with the observed negative CP of Sgr A* to constrain the most plausible magnetic field configuration.
\section{Polarized Radiative Transfer and CP Mechanism}
\label{sec:cp_mechanism}

The evolution of polarized radiation through a relativistic, magnetized plasma
is governed by the time independent radiative transfer equation. Along a ray
parameterized by the affine parameter $s$, the Stokes vector
$\bm{S} = (I, Q, U, V)^{\rm T}$ evolves as:
\begin{equation}\label{eq:GenRadTrans}
\frac{d}{ds}
\begin{pmatrix} I \\ Q \\ U \\ V \end{pmatrix}
=
\begin{pmatrix} j_I \\ j_Q \\ j_U \\ j_V \end{pmatrix}
-
\begin{pmatrix}
\alpha_I  & \alpha_Q & \alpha_U & \alpha_V \\
\alpha_Q  & \alpha_I & \rho_V   & -\rho_U  \\
\alpha_U  & -\rho_V  & \alpha_I & \rho_Q   \\
\alpha_V  & \rho_U   & -\rho_Q  & \alpha_I
\end{pmatrix}
\begin{pmatrix} I \\ Q \\ U \\ V \end{pmatrix},
\end{equation}
where $j_S$, $\alpha_S$, and $\rho_S$ denote the emission, absorption, and
Faraday rotation and conversion coefficients for each Stokes component $S \in (I, Q, U, V)$,
respectively. We follow the IEEE convention in which $V > 0$ corresponds to
right handed circular polarization \cite{Hamaker:1996stokes}.
Isolating the equation for Stokes $\rm V$:
\begin{equation}\label{eq:StokesV}
\frac{d}{ds}V = j_V - \alpha_V I - \rho_U Q + \rho_Q U - \alpha_I V,
\end{equation}
this equation identifies two mechanisms that can generate circular polarization during propagation.

The first mechanism is intrinsic circularly polarized synchrotron emission,
$j_V$. For a thermal (Maxwell--J\"{u}ttner) distribution, the intrinsic emissivity is \cite{Dexter:2016grtrans}:
\begin{equation}\label{eq:jV}
j_V(\nu, \theta) = \frac{2 n_{\rm e} {\rm e}^2 \nu \cot\theta}{3\sqrt{3}\,c\,\Theta_{\rm e}^2}
\,I_V(x),
\end{equation}
where 
\begin{equation}
\begin{split}
I_V(x)=&
\left(
1.8138x^{-1}
+3.423x^{-2/3}
+0.02955x^{-1/2}
\right. \\
&\left.
+2.0377x^{-1/3}
\right)
\exp\left(-1.8899x^{1/3}\right),
\end{split}
\end{equation}
$\theta$ is the angle between the photon wavevector $\vec{k}$ and magnetic field $\vec{B}$, $\Theta_{\rm e} \equiv k_{\rm B} T_{\rm e} / m_{\rm e} c^2$ is the dimensionless electron temperature.
$n_{\rm e}$ is the electron number density, $m_{\rm e}$, $T_{\rm e}$ and ${\rm e}$  are the electron mass, the electron temperature and elementary charge, respectively. $\nu$ is the emitted frequency, which can be inferred from the observed frequency $\nu_{\rm obs}$ with the redshift and Doppler effect. $x = \nu / \nu_c$  and the characteristic synchrotron frequency $\nu_c = \frac{3}{2}\nu_{\rm B} \sin\theta\ \Theta_{\rm e}^2$, where the electron cyclotron frequency $\nu_{\rm B} ={\rm e} |B| / (2\pi m_{\rm e} c)$ and $|B|$ is the magnetic field magnitude. The sign of intrinsically emitted circular polarization directly encodes the direction of the magnetic field with respect to the photon wavevector.

The second mechanism is Faraday conversion, $\rho_Q$. The conversion coefficient is \cite{Dexter:2016grtrans}:
\begin{equation}\label{eq:rhoQ}
\rho_Q = \frac{n_{\rm e} {\rm e}^2 \nu_{\rm B}^2 \sin^2\theta}{m_{\rm e} c \nu^3}\,f(X)
\left[\frac{K_1(\Theta_{\rm e}^{-1})}{K_2(\Theta_{\rm e}^{-1})} + 6\Theta_{\rm e}\right],
\end{equation}
where 
\begin{equation}\label{eq:X}
    X=\left(\frac{3}{2 \sqrt{2}} 10^{-3} \frac{\nu}{\nu_{c}}\right)^{-1/2},
\end{equation}
\begin{equation}
\begin{split}
f(X)=&
2.011\exp\left(-\frac{X^{1.035}}{4.7}\right) \\
&-\cos\left(\frac{X}{2}\right)
\exp\left(-\frac{X^{1/2}}{2.73}\right) \\
&-0.011\exp\left(-\frac{X}{47.2}\right),
\end{split}
\end{equation}
and $K_1$, $K_2$ are modified Bessel functions of the second kind. In a field-aligned Stokes basis $j_U = \alpha_U = \rho_U = 0$, and the Faraday conversion term in the transfer equation reduces to $+\rho_Q U$. Stokes U is therefore required to produce Stokes V by Faraday conversion, whereas synchrotron emission initially produces linear polarization entirely in the Stokes $Q$ component (i.e., $U = 0$). For Faraday conversion to operate, a physical mechanism must first transfer linear polarization from $Q$ into $U$. This is typically achieved through two pathways. The first is Faraday rotation ($\rho_V$) \cite{Dexter:2016grtrans}, 
\begin{equation}\label{eq:rhoV}
\rho_V = \frac{2 n_{\rm e} {\rm e}^2 \nu_{\rm B}}{m_{\rm e} c \nu^2}
\frac{K_0(\Theta_{\rm e}^{-1})}{K_2(\Theta_{\rm e}^{-1})}\cos\theta g(X),
\end{equation}
where
\begin{equation}
    g(X) = 1 - 0.11 \ln{(1 + 0.035 X)},
\end{equation}
which can change the plane of polarization as the radiation propagates by directly exchanging $Q$ and $U$ \cite{Ricarte:2021cpmag}. The other way relies on the geometric twisting of the magnetic field along the line of sight. As the magnetic field twists along the photon trajectory, it rotates the polarization basis, thereby projecting the initial $Q$ component into $U$ \cite{Ricarte:2021cpmag}.

An important consideration when comparing models to observations is that
$V_{\rm net}$ constrains the net handedness of the integrated emission but
does not directly determine the absolute polarity of the magnetic field
threading the accretion flow. In a semi-analytic model, the field polarity is a free parameter set by the sign convention of the vector potential, and
both orientations must be explored. We therefore apply the field reversal
$\vec{B} \rightarrow -\vec{B}$ systematically to each configuration, and compare the corresponding results between the aligned ($\vec{B}$) and reversed field ($-\vec{B}$) distributions. In addition, we can understand  whether each model's CP is produced via ($\vec{B}$) field polarity invariant pathways (Faraday conversion through field twist) or non-invariant pathways (intrinsic emission or Faraday conversion through Faraday rotation) \cite{Joshi:2024bth}.

To connect our synthetic images to observations of Sgr A*, we compute two metrics. The net circular polarization fraction,
\begin{equation}\label{eq:vnet}
    V_{\rm net} \equiv \frac{\int d^2x\, V(x,y)}{\int d^2x\, I(x,y)}
\end{equation}
is the image integrated CP fraction accessible to spatially unresolved ALMA observations. The average absolute circular polarization fraction,
\begin{equation}\label{eq:vavg}
    \langle | V | \rangle \equiv \frac{\int d^2x\, |V|}{\int d^2x\, I}.
\end{equation}
sums the absolute value of the local CP fraction across the image, capturing the spatially resolved CP structure.
\section{Model and Numerical Setup}
\label{sec:mns}
We investigate the circular polarization properties of Sgr A* using a stationary axisymmetric Radiatively Inefficient Accretion Flow (RIAF) framework in Kerr spacetime \cite{Kerr:1963metric,Pu:2018sgrageometry,Pu:2016dynamics,Wang:2025btn}. We specify the electron number density $n_{\rm e}$ and electron temperature $T_{\rm e}$ as:
\begin{align}
n_{\rm e} &= n_{\rm e,0} \left( \frac{r}{r_g} \right)^{-\delta} \exp \left[- \frac{1}{2} (H \tan \theta)^{-2} \right], \label{eq:ne} \\ 
T_{\rm e} &= T_{\rm e,0} \left( \frac{r}{r_g} \right)^{-\gamma} , \label{eq:Te}
\end{align}
where $r_g = GM_{\rm BH}/c^2$ is the gravitational radius. The parameter $H$ controls the geometric thickness of the disk \cite{Vincent:2022photonring,Saurabh:2025m87semi,Zhang:2024jrw}, we adopt $H=0.5$ as our fiducial model. For the temperature profile, we set $\gamma=0.84$, a value consistent with observational constraints from radiative spectra \cite{Broderick:2011lowspin}. To describe the kinematics of the accretion flow, we model the four velocity by interpolating between Keplerian orbital motion and geodesic free fall, following the prescription of \cite{Saurabh:2025m87semi,Yin:2025rao}. The velocity field is governed by two parameters, $\kappa_{\rm K}$ and $\kappa_{\rm ff}$, which control the weights of the Keplerian and radial infall components, respectively.
\begin{align}
    u^r &= u^r_{\rm K} + \kappa_{\rm ff}(u^r_{\rm ff} - u^r_{\rm K}),\\
    \Omega &= \Omega_{\rm K} + (1-\kappa_{\rm K})(\Omega_{\rm ff} - \Omega_{\rm K}),
\end{align}
where $\Omega = u^\phi/u^t$ is the angular velocity. We adopt $(\kappa_{\rm ff},\kappa_{\rm K}) = (0.5,0.5)$ as our fiducial model. The relevant model parameters are summarized in Table~\ref{table:param}.

\begin{table} 
    \centering
    \begin{tabular}{c c c}
        \hline
        Parameter & Value & Parameter Description\\
        \hline
        $M_{\rm BH}$ & $4.3\times10^6\rm{M}_\odot$ & Black hole mass \\
        $D_{s}$ &  $8.3 \times 10^3 \rm{pc}$ & Distance to the source\\
        $\delta$ & 1.1 & $n_{\rm e}$ power law index\\
        $\gamma$ & 0.84 & $T_{\rm e}$ power law index\\
        $H$ & $0.5$ & Disk thickness\\
        $a$ & $0.5$ & Black hole spin\\
        $i$ & $30^\circ$ & Inclination angle\\
        $\kappa_{\rm K}$ & $0.5$ & Keplerian parameter\\
        $\kappa_{\rm ff}$ & $0.5$ & Radial infall parameter\\
        $B_0$ & $29$ G & Magnetic field strength\\
        $n_{\rm e,0}$ & $\displaystyle 10^{6}\text{--}10^{7}\,\rm{cm}^{-3}$ & The normalization of\\
         &  &  electron distribution\\
        $T_{\rm e,0}$ & $5\times10^{11}\,\rm{K}$ & The normalization of temperature\\
        $\nu_{\rm obs}$ & $230$\,GHz & Observing frequency\\
        $p_0$ & -100  & Combined field parameter\\
        FOV & $200\,\mu\rm{as}$ & Field of view\\
        $N_X \times N_Y$ & $200 \times 200\,\,\rm{px}$  & Number of pixels\\
        \hline
    \end{tabular}
    \caption{Fiducial Parameters for RIAF}
    \label{table:param}
\end{table}
    
Embedded within this accretion flow, we introduce distinct magnetic field geometries to study their imprint on the circular polarization. These configurations are defined primarily through the vector potential component $A_\phi$, assuming axisymmetry ($A_r = A_\theta = 0$). This approach enforces the divergence-free condition $\nabla_i B^i = 0$ by construction.

\begin{figure*}[htbp]
\centering
\includegraphics[width=\textwidth]{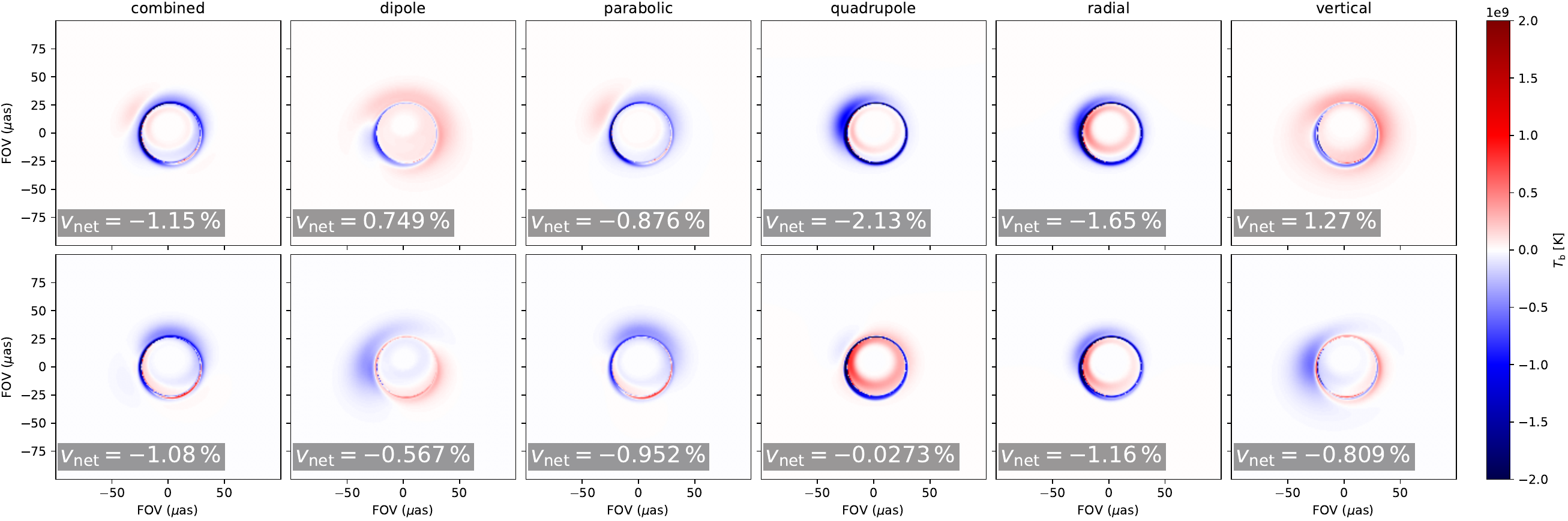}
\caption{The images of Stokes $\rm V$ for the fiducial parameters with various magnetic field configurations. The upper (lower) row shows results for the aligned (reversed) field configurations.}
\label{fig:cp_maps}
\end{figure*}

We note that the frame-dragging effect in the Kerr metric twists magnetic field lines, inevitably inducing an electric field. As demonstrated by \cite{Komissarov:2004electrodynamics} and \cite{Komissarov:2022charged}, such induced electric fields cannot be completely screened in the vicinity of a rotating black hole, particularly within the ergosphere. While this is a general feature of semi-analytic models where the field is not evolved via the full induction equation, we focus here specifically on the impact of the resulting magnetic field geometry on the polarimetric observables. The MAD state, currently identified as the best-bet model for Sgr A* based on EHT and multi-wavelength constraints \cite{EHTSgrA-V}, exhibits strong poloidal magnetic field components, motivating our focus on poloidal magnetic geometries. However, within the MAD framework, the precise geometry of the poloidal magnetic field near the horizon remains poorly constrained. Here, as in Ref.\cite{Saurabh:2025m87semi}, we consider the following six geometries of poloidal magnetic fields:
\begin{enumerate}
    \item Radial (split monopole) \cite{Blandford:1977bz}:
         \begin{equation}
         A_{\phi, R} =  1 - |\cos{\theta}|.
        \end{equation}
    \item Vertical \cite{Vos:2022yij}:
         \begin{equation}
         A_{\phi, V} =  r \sin{\theta}.
        \end{equation}
    \item Dipole \cite{McKinney:2009stability}:
        \begin{equation}
        A_{\phi, D} =  \frac{1}{2}\bigg[(r + r_0)^\nu\mathcal{F}_{-} + 2 \mathcal{F}_+(1 - \ln\mathcal{F}_+)\bigg],
        \end{equation}
      where $\mathcal{F}_\pm = 1 \pm \cos^\mu{\theta}$, with parameters $\nu = 3/4$, $\mu = 4$, and $r_0=4r_g$. In the equatorial plane, this field is predominantly vertical.
    \item Quadrupole \cite{McKinney:2009stability}:
        \begin{equation}
        A_{\phi, Q} = A_{\phi, D} \cos{\theta}.
        \end{equation}
    \item Parabolic \cite{Tchekhovskoy:2010radioloud}:
        \begin{equation}
        A_{\phi, P} = r^k (1 - |\cos{\theta}|) = r^k A_{\phi, R},
        \end{equation}
        where we set $k=0.75$. For $k=0$, this expression reduces to the radial (split monopole) field.
    \item Combined Field \cite{Kenzhebayeva:2024combined}: A self-consistent solution to Maxwell's equations in Kerr spacetime combining the vertical homogeneous solution \cite{Wald:1974uniform} and the radial split-monopole:
        \begin{equation}
       A_{\phi, C} = g_{\phi\phi} + 2 a g_{t\phi} + p_0\frac{(r^2 + 
       a^2)|\,\cos{\theta}\,|}{r^2 + a^2 \cos^2{\theta}}.  \label{eq:combined}
    \end{equation}
    Here, $p_0$ controls the ratio of the radial to vertical magnetic field components. This field asymptotes to a constant vertical field at large distances.
\end{enumerate}
The above six magnetic field configurations considered here span the range of poloidal topologies commonly invoked in models of black hole accretion, enabling a systematic study of how polarimetric observables depend on field geometry.

For the poloidal configurations defined via the vector potential, the magnetic field components in the static observer's frame (lab-frame) are derived as:
 \begin{equation}
           B^r =  \frac{\partial_\theta A_\phi }{\sqrt{-g}}\ , \ \ B^\theta =  -\frac{\partial_r A_\phi }{\sqrt{-g}}\ , \ \ B^\phi = B^t = 0 .
\end{equation}
we renormalize the magnetic fields by a constant factor such that the field strength at the equatorial ring $(r = 3 r_g, \theta = \pi/2)$ equals $B_0 \approx 30$\,G, which is comparable to the estimate derived from EHT \cite{EHTSgrA-V}. We transform the lab-frame magnetic field $B^\mu$ to the fluid comoving frame four vector $b^\mu$ following the ideal MHD relation \cite{Gammie:2003harm}:
 \begin{equation}
           b^t = B^iu_i \ \  , \ \ b^i = \frac{B^i + b^tu^i}{u^t} ,
\end{equation}
where $u^\mu$ is the fluid four-velocity defined previously. 

We synthesize polarized images by solving the general relativistic radiative transfer (GRRT) equations using the numerical code \cite{Marszewski:2021ipole}. For the physical parameters of Sgr A*, we adopt the black hole mass and distance measurements derived from the EHT and GRAVITY collaborations: $M_{\rm BH} = 4.3\times10^6 \rm{M}_\odot$ and $D_{s} = 8.3\,\rm{kpc}$, respectively \cite{GRAVITY:2018redshift,EHTSgrA-IV}. We set the observing frequency to $\nu_{\rm obs} = 230\,$GHz \cite{EHTSgrA-II}. Following the "best-bet" models from the EHT and multi-wavelength constraints which favor a low-inclination geometry, we adopt a fiducial inclination angle of $i = 30^\circ$, and the black hole spin is 0.5 \cite{EHTSgrA-V}. Unlike M87*, Sgr A* lacks a definitive large-scale jet to constrain the position angle; thus, for simplicity, we align the position angle of the BH spin vector to the vertical direction ($0^\circ$ East of North) on the observer's screen. We ray trace images with a field of view (FOV) of $200\,\mu$as, which amply covers the photon ring and the extended emission region. For each magnetic field model and accretion flow setup, we vary the normalization of electron number density scale $n_{\rm e,0}$ and take $T_{e,0} = 5\times10^{11} K$ to match a total flux density of $\mathcal{I}_{\rm tot} \approx 2.4$\,Jy at 230\,GHz, consistent with ALMA measurement \cite{Wielgus:2022a}.    
\section{Result}\label{sec:res}
We now present the circular polarization properties of our RIAF models. We begin by examining the CP morphology across the six magnetic field geometries for our fiducial parameters (Section~\ref{subsec:cp_geometry}). We then perform a coefficient suppression analysis to identify the dominant CP production mechanisms (Section~\ref{subsec:coeff}). Finally, we investigate the dependence of $V_{\rm net}$ on black hole spin (Section~\ref{subsec:spin}) and observer inclination (Section~\ref{subsec:inclination}).
\begin{figure*}[htbp]
\centering
\includegraphics[width=\textwidth]{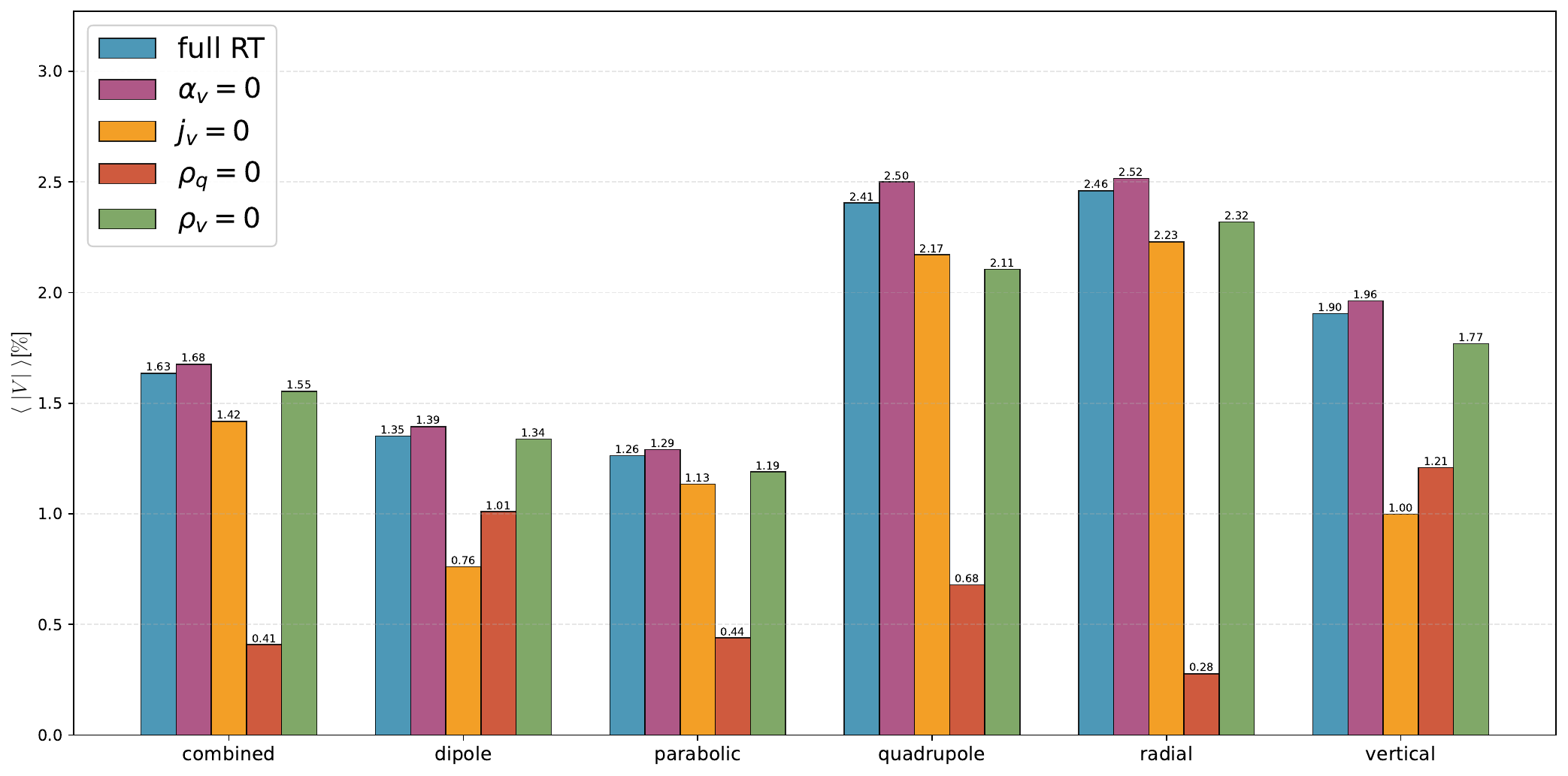}
\caption{The resolved CP fraction $\langle|V|\rangle$ for the fiducial parameters with various magnetic field geometries under full radiative transfer (blue) and with individual coefficients suppressed: $\alpha_V = 0$ (purple), $j_V = 0$ (orange), $\rho_Q = 0$ (red), and $\rho_V = 0$ (green).}
\label{fig:coeff_suppression}
\end{figure*}
\subsection{CP with Different Magnetic Field Geometry}
\label{subsec:cp_geometry}
Figure~\ref{fig:cp_maps} presents the images of Stokes $\rm V$ with various magnetic field geometries for the fiducial model. The bright positive feature is the region of the accretion disk where the fluid velocity is aligned with the line of sight. Its opposite sign is a consequence of the Faraday thin of the image and the imprint of the magnetic fields on Stokes $\rm V$ through Faraday conversion as observed in \cite{Ricarte:2021cpmag}. Upon field reversal, the combined, parabolic, and radial configurations maintain the same sign of $V_{\rm net}$, indicating that the CP sign is insensitive to the magnetic field polarity. In contrast, the dipole and vertical configurations are sensitive to field polarity, their $V_{\rm net}$ inverts sign upon reversal.
\subsection{Contribution of Transfer Coefficients}
\label{subsec:coeff}
To diagnose the relative contributions of the two CP production mechanisms identified in Section~\ref{sec:cp_mechanism}, we perform radiative transfer coefficients suppression analysis: for each magnetic field configuration, we repeat the full radiative transfer calculation four times, setting $\alpha_V = 0$ (CP absorption), $j_V = 0$ (intrinsic CP emission), $\rho_Q = 0$ (Faraday conversion) and $\rho_V = 0$ (Faraday rotation) in turn, and compare the resulting $\langle|V|\rangle$ against the full radiative transfer (full RT). The results are presented in Figure~\ref{fig:coeff_suppression}.

Across all six configurations, CP absorption ($\alpha_V$) plays a negligible role. This is because the models are optically thin. For the combined, parabolic, quadrupole, and radial configurations, $\rho_Q$ the dominant mechanism of CP production, as $\langle|V|\rangle$ is highly suppressed when setting $\rho_Q = 0$. Although $j_V$ is sub-dominant, it is non-negligible. The dipole and vertical configurations show a different response. Setting $j_V=0$ produces the larger reduction $44\%$ for the dipole field and $47\%$ for the vertical one, while suppressing $\rho_Q$ yields a smaller but non-negligible decrease of $25\%$ and $36\%$, respectively. Intrinsic CP emission is therefore the primary mechanism in these two geometries, with Faraday conversion playing a secondary role.
\subsection{Effect of Black Hole Spin on The Net CP}
\label{subsec:spin}
\begin{figure*}[p]
\centering
\includegraphics[width=0.6\textwidth]{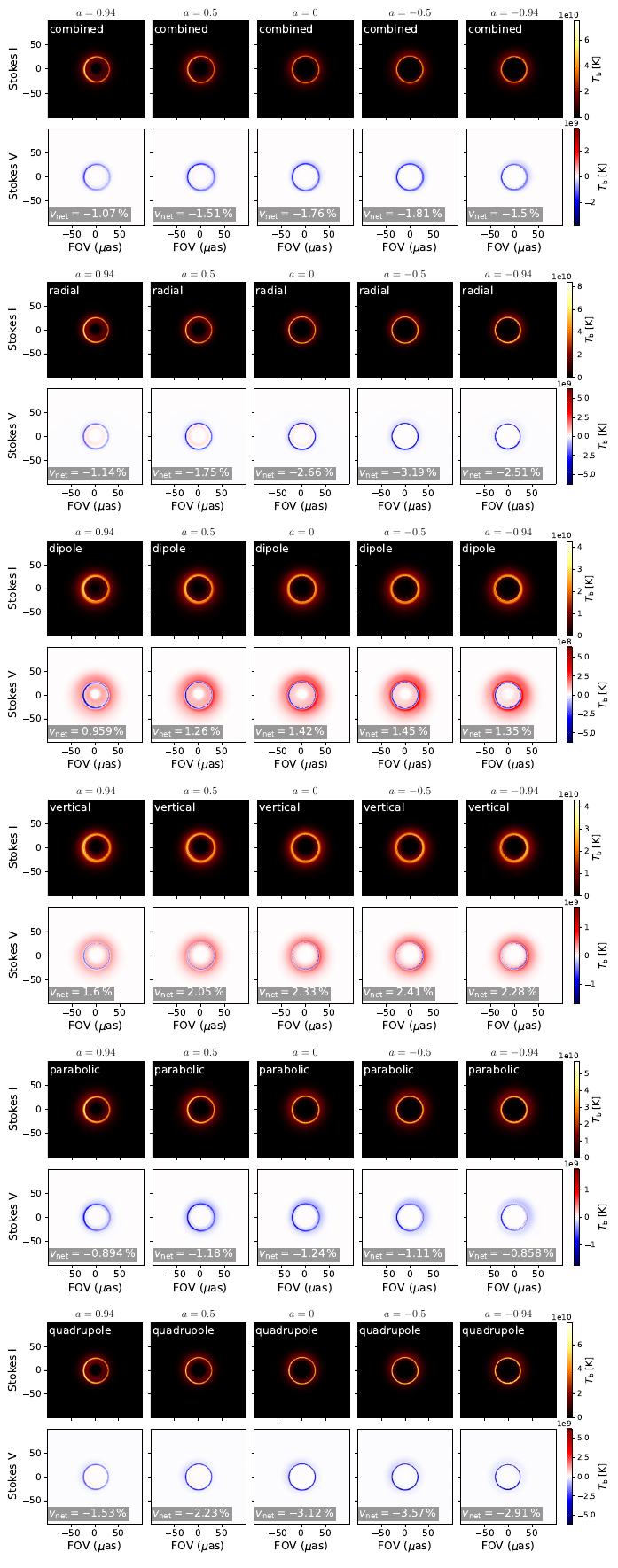}
\caption{The images of Stokes $\rm I$ and $\rm V$ across spin for various magnetic field geometries, inclination $10^\circ$, disk thickness 0.5.}
\label{fig:spin}
\end{figure*}

We investigate the dependence of $V_{\rm net}$ on black hole spin. Figure~\ref{fig:spin} presents the Stokes $\rm I$ and $\rm V$ images for the all magnetic field configurations with five representative spin values. It shows that as the spin parameter ranges from $a = -0.94$ to 0.94, the absolute value $|V_{\rm net}|$ first increases and then decreases for each magnetic field configuration.
Moreover, for each spin, a key morphological distinction in the Stokes $\rm V$ for different field geometries lies in 
the relative sign of the $n=1$ photon ring with respect to the $n=0$ direct emission. For the dipole and vertical field configurations, the $n=1$ photon ring carries the opposite sign of circular polarization compared to the $n=0$ component. By contrast, for the radial, quadrupole, combined, and parabolic field geometries, the $n=1$ photon ring and the $n=0$ component share the same sign of Stokes $\rm V$. As the accretion disk is prograde with respect to the black hole spin (i.e., $a>0$), $|V_{\rm net}|$ across all six field geometries is found to be lower at high spin case, while  the situation is more complicated in the retrograde case.

\begin{figure*}[htbp]
\centering
\includegraphics[width=\textwidth]{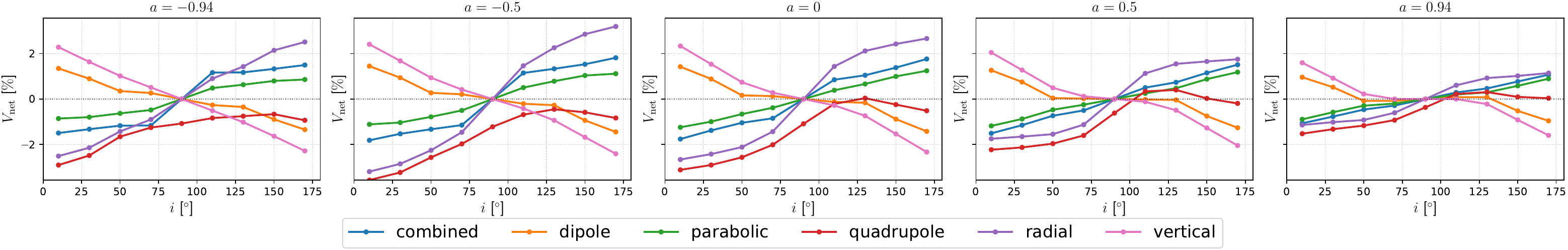}
\caption{Dependence of net CP $V_{\rm net}$ on the inclination for various magnetic field geometry, disk thickness 0.5}
\label{fig:inclination}
\end{figure*}

\subsection{Effect of Inclination on The Net CP}
\label{subsec:inclination}
\begin{figure*}[htbp]
\centering
\includegraphics[width=\textwidth]{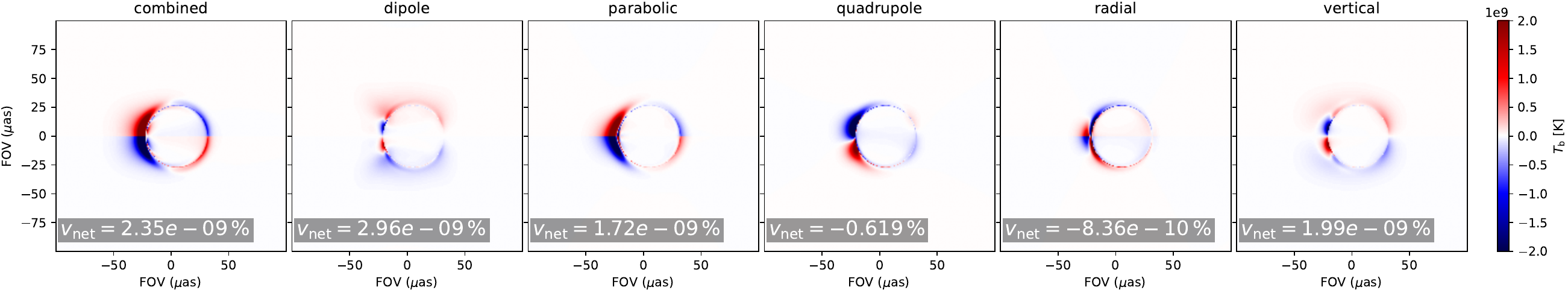}
\caption{The image of Stokes V with various magnetic field geometries for spin +0.5, disk thickness 0.5, inclination $90^\circ$.}
\label{fig:edgeon}
\end{figure*}
Figure~\ref{fig:inclination} shows effect of inclination on the Net CP for six magnetic field geometries. The net circular polarization $V_{\rm net}$ decreases with observer inclination in dipole and vertical magnetic field configurations, while it exhibits an increasing trend in the combined, parabolic, and radial geometries. In the quadrupole magnetic field case, $V_{\rm net}$  first increases and subsequently decreases. Generally,
images at observer inclinations $\theta_i \neq 90^\circ$ typically contain contributions from both the near-side and far-side regions. For both of these regions, the intrinsic emission related to the quantity $\bm{\vec{k}} \cdot \bm{\vec{B}}$ and the geometric twisting of the magnetic field along the line of sight are important.
Figure~\ref{fig:inclination} shows that for $\theta < 90^\circ$ in the aligned magnetic field configuration, the dipole and vertical magnetic field geometries yield a positive $V_{\rm net}$. The main reason is that in the near-side region the intrinsic emission $j_V > 0$ and $\rho_V > 0$
because the total quantity $\bm{\vec{k}} \cdot \bm{\vec{B}} > 0$.  In addition, $\rho_V > 0$ in the Faraday thin regime
enhances Stokes $\rm U$ and subsequently increases Stokes $\rm V$ through Faraday conversion \cite{Ricarte:2021cpmag}.
Furthermore, the radial, quadrupole, parabolic, and combined magnetic field geometries which are dominated by Faraday conversion through the magnetic field twist exhibit a negative $V_{\rm net}$ which is driven by the counter-clockwise vertical twist of the magnetic field along the line of sight. This counter-clockwise twist initially generates a negative Stokes $\rm U$ component ($\rm U < 0$), which Faraday conversion subsequently and naturally transforms into a negative CP signal. Furthermore, viewing the system from a complementary inclination angle reverses both the wavevector $\bm{\vec{k}}$ and the direction of the magnetic field twist, which consequently inverts the sign of the CP.

Figure~\ref{fig:edgeon} show that the symmetry of the global magnetic field across the midplane for the radial, vertical, dipole, parabolic, and combined magnetic field geometries, any additional twist effects cancel each other out as $\theta \to 90^\circ$ (edge-on view), resulting in $V_{\rm net} \approx 0$. In contrast, the quadrupole field uniquely produces a non-zero $V_{\rm net}$. This is because the radial component of the quadrupole field maintains the same direction both above and below the disk and the geometric line of sight magnetic twist does not flip its sign across the midplane. As a result, Faraday conversion produces Stokes $\rm V$ of the same sign in both hemispheres. These contributions add constructively rather than destructively, yielding a distinctly non-zero net circular polarization for the edge-on quadrupole system.

\begin{figure*}[htbp]
\centering
\includegraphics[width=\textwidth]{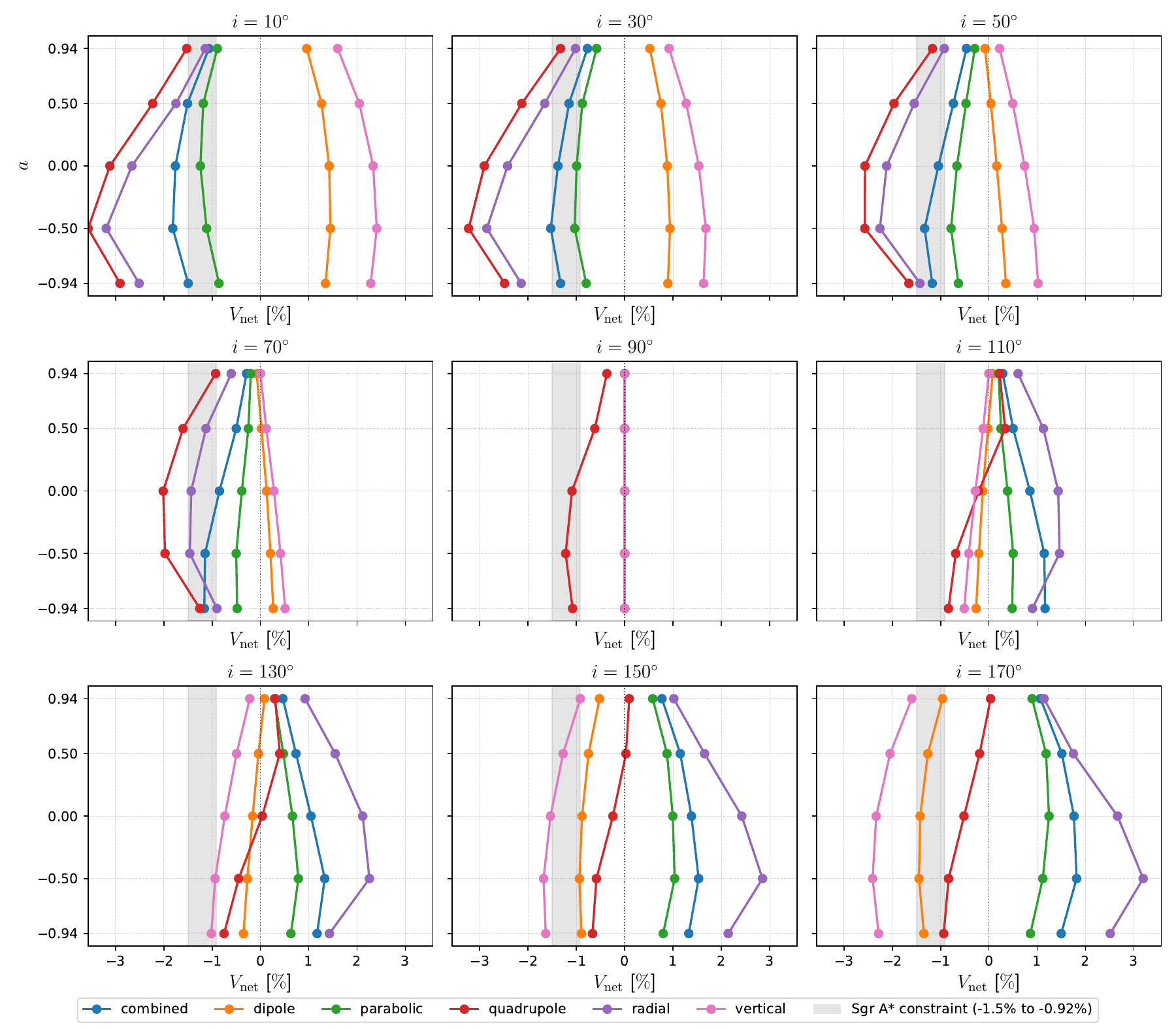}
\caption{Dependence of net circular polarization fraction $V_{\rm net}$ on magnetic field geometry and black hole spin $a$ with different inclinations for the aligned field configuration. The gray shaded band indicates the observational constraint from ALMA measurements of Sgr A*, $V_{\rm net} \in [-1.5\%, -0.92\%]$.}
\label{fig:vnet_aligned}
\end{figure*}

\begin{figure*}[htbp]
\centering
\includegraphics[width=\textwidth]{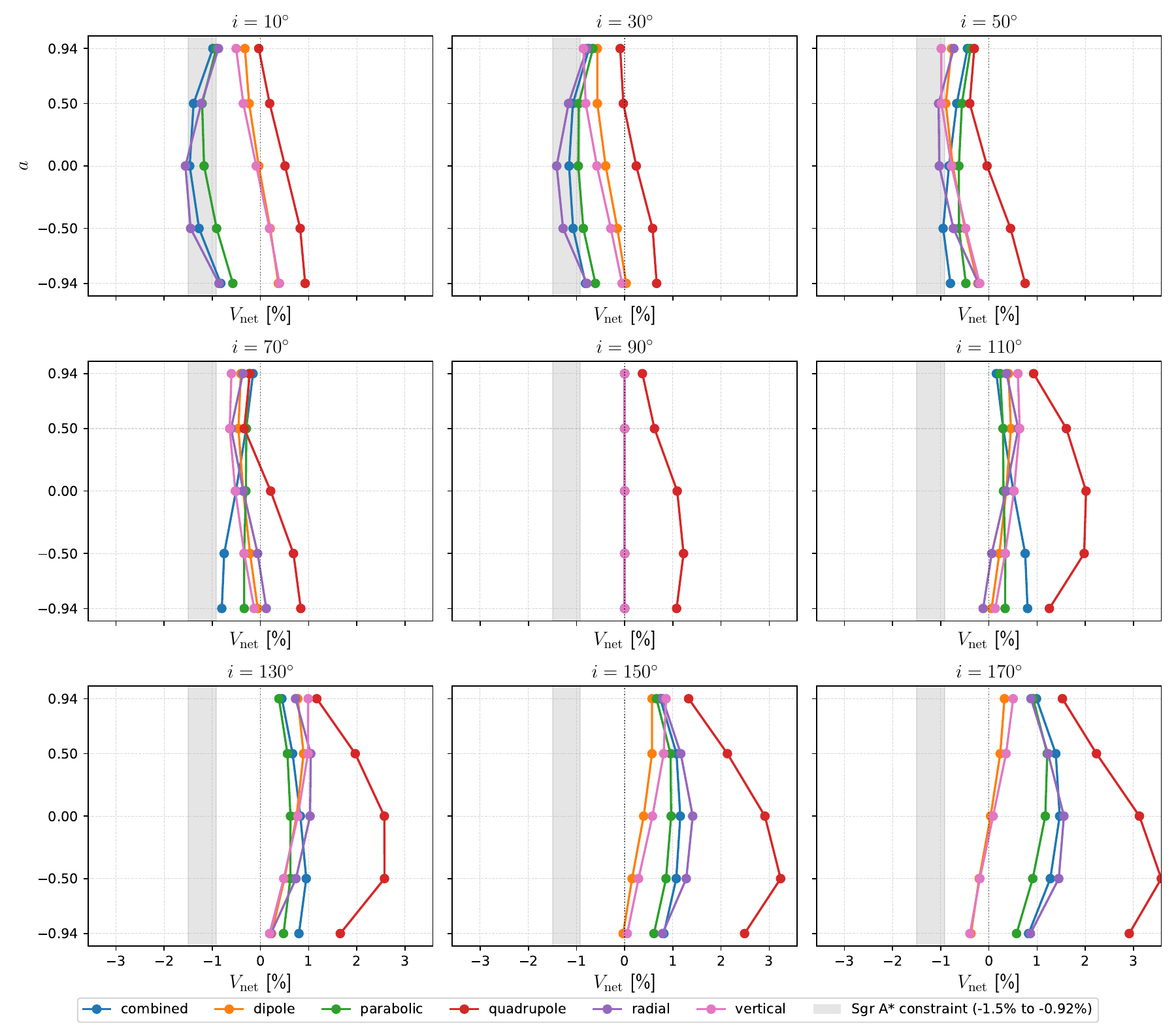}
\caption{Same as Figure~\ref{fig:vnet_aligned} but for the 
reversed field configuration.}
\label{fig:vnet_reversed}
\end{figure*}
\section{Observational Constraints}
\label{sec:obs}
To assess which magnetic field geometries are compatible with Sgr A*, we compare the predicted net circular polarization fraction $V_{\rm net}$ from our RIAF models against the persistent ALMA measurements at 230\,GHz. The observational constraint is drawn from 8 data points, reported in \cite{Bower:2018interferometric,Wielgus:2022alma}, which span $V_{\rm net}$ values between $-1.5\%$ and $-0.92\%$.

Figures~\ref{fig:vnet_aligned} and \ref{fig:vnet_reversed} illustrate the dependence of $V_{\rm net}$ on black hole spin $a$ across six magnetic field geometries and nine observer inclinations ($10^\circ$ to $170^\circ$) for the aligned and reversed field polarities, respectively. These plots reveal several distinct systematic trends.

 No magnetic field geometry produces consistent negative $V_{\rm net}$ at all inclinations and spins. The sign and magnitude of $V_{\rm net}$ depend sensitively on the combination of field geometry, polarity, spin, and inclination, confirming that the ALMA CP constraint is a powerful discriminant of magnetic field architecture. For the reversed field configuration, no magnetic field geometry produces $V_{\rm net}$ within the observed constraint band at high observer inclinations $i > 90^\circ$.The radial and parabolic field configurations exhibit qualitatively distinct behavior under polarity reversal: unlike the dipole, vertical, or combined geometries, their $V_{\rm net}$ values do not undergo a sign flip upon $\vec{B} \rightarrow -\vec{B}$, but instead shift modestly in magnitude while retaining the same sign across a wide range of inclinations and spins. For virtually all field geometries and both polarities, models at $i = 90^\circ$ produce $V_{\rm net} \approx 0$, failing the Sgr A* constraint. 
\section{conclusions}
\label{sec:con}
In this paper, we investigate the circular polarization of simulated images of Sgr A* with various magnetic field geometries, using a stationary semi-analytic RIAF model, focusing on the net circular polarization fraction ($V_{\rm net}$). We explore models that cover different black hole spins, observer inclinations, and the alignment and reversal polarities of the global magnetic field, in order to understand the characteristics of $V_{\rm net}$ under different magnetic field geometries. We conducted a radiative transfer coefficients suppression analysis, followed by plotting the $V_{\rm net}$ distribution and comparing it with unresolved ALMA observations of Sgr A*. We find the following results. 

\begin{itemize}
    \item Field reversal does not simply invert the $V_{\rm{net}}$ distribution due to the interplay of symmetric and antisymmetric terms in the polarized radiative transfer equation. Crucially, the dominant term depends on the magnetic field geometry: some configurations are governed primarily by symmetric processes, while others are driven by antisymmetric emission, with most models exhibiting a complex mixture of both.
    \item The spin of a black hole affects the $V_{\rm{net}}$ distribution of the model. As the spin parameter ranges from $a = -0.94$ to 0.94, the absolute value $|V_{\rm net}|$ first increases and then decreases for each magnetic field configuration.
    As the accretion disk is prograde with respect to the black hole spin, the CP production across all six field geometries is found to be lower at high spin case, while the situation is more complicated in the retrograde case.  Moreover, the net CP observed from edge-on views $V_{\rm net} \approx 0$ except for the quadrupole geometry.
    \item For Face-on models $(i<90^\circ)$, dipolar and vertical magnetic field geometries produce positive $V_{\rm{net}}$ encoding an overall clockwise twist of field, whereas combined, parabolic, and radial magnetic field geometries produce negative $V_{\rm{net}}$ encoding an overall counter-clockwise twist of field.
    \item In edge-on models $(i=90^\circ)$, quadrupolar magnetic field geometries produce a non-zero net circular polarization $V_{\rm{net}} \neq 0$, whereas other field configurations yield $V_{\rm{net}} \approx 0$, due to symmetries in the magnetic field structure.
    \item For all magnetic field geometries with reversed fields, models at high inclination angles are inconsistent with observational constraints, as they produce positive values of $V_{\text{net}}$.
\end{itemize}

We emphasize that  circular polarization images  provide  constraints on magnetic field structure complementary to those from linear polarization images. The recent studies \cite{Saurabh:2025m87semi} of linear polarization images by the semi-analytic RIAF  for various magnetic field geometries reveal that toroidal-dominated fields produce radial EVPA patterns and poloidal-dominated fields yield azimuthal patterns. 
The CP analysis here complements these LP studies by probing magnetic field polarity reversals along the line of sight information inaccessible through LP alone. For example, a dipole field and its polarity-reversed counterpart produce identical EVPA patterns but opposite $V_{\rm net}$ signs. Moreover, different poloidal field configurations can be distinguished reliably by comparing the CP signals, providing additional constraints beyond LP observations.

\newpage
Overall, our analysis reveals that the net circular polarization of 
Sgr A* is governed by the interplay between magnetic field geometry, 
black hole spin, and observer inclination, with polarity-invariant 
(Faraday conversion dominated) and polarity-sensitive (emission dominated) 
geometries exhibiting distinct observational signatures. The persistent 
negative CP ($V_{\rm net} \in [-1.5\%, -0.92\%]$) excludes reversed-field 
configurations at high inclinations. These results establish circular polarization 
as a decisive diagnostic for constraining the magnetic field geometry 
of Sgr A*.
\begin{acknowledgments}
This work was supported by the National Natural Science Foundation of China under Grant No.12275078, 11875026, 12035005, 2020YFC2201400 and the innovative research group of Hunan Province under Grant No. 2024JJ1006.
\end{acknowledgments}
\noindent\textbf{Conflict of Interest}\quad The authors declare that they have no conflict of interest.
\bibliography{bz_bbh}
\bibliographystyle{sn-aps}
\end{document}